\documentclass{ws-procs9x6}
\usepackage[svgnames]{xcolor}
\usepackage[colorlinks,allcolors=MidnightBlue]{hyperref}

\begin{document}
{\noindent\footnotesize Forthcoming in: {\it Probing the Meaning of Quantum Mechanics}. Aerts, D. \& Arenhart, J. R. B. \& de Ronde, C. \& Sergioli, G. (eds.) World Scientific, 2023. URL: \url{www.worldscientific.com/worldscibooks/10.1142/13602}.}

\title{THE ELIMINATION OF METAPHYSICS THROUGH THE EPISTEMOLOGICAL ANALYSIS:\\LESSONS (UN)LEARNED FROM METAPHYSICAL UNDERDETERMINATION}

\author{RAONI W. ARROYO$^*$}
\address{Department of Philosophy, Communication and Performing Arts\\Roma Tre University\\Rome, Italy\\Centre for Logic, Epistemology and the History of Science\\University of Campinas\\Fellow Researcher of the S\~ao Paulo Research Foundation (FAPESP)\\Research Group in Logic and Foundations of Science (CNPq)\\International Network on Foundations of Quantum Mechanics and Quantum Information\\$^*$E-mail: \url{rwarroyo@unicamp.br}}

\author{JONAS R. BECKER ARENHART}
\address{Department of Philosophy\\Federal University of Santa Catarina\\Florian\'opolis, Santa Catarina, Brazil\\Graduate Program in Philosophy\\Federal University of Maranh\~ao\\Research Group in Logic and Foundations of Science (CNPq)\\International Network on Foundations of Quantum Mechanics and Quantum Information\\E-mail: \url{jonas.becker2@gmail.com}}

\author{D\'ECIO KRAUSE}
\address{Graduate Program in Logic and Metaphysics\\Federal University of Rio de Janeiro\\Rio de Janeiro, Brazil\\Research Group in Logic and Foundations of Science (CNPq)\\International Network on Foundations of Quantum Mechanics and Quantum Information\\E-mail: \url{deciokrause@gmail.com}}

\begin{abstract}
Abstract: This chapter argues that the general philosophy of science should learn metaphilosophical lessons from the case of metaphysical underdetermination, as it occurs in non-relativistic quantum mechanics. Section \ref{sec:2} presents the traditional discussion of metaphysical underdetermination regarding the individuality and non-individuality of quantum particles. Section \ref{sec:3} discusses three reactions to it found in the literature: eliminativism about individuality; conservatism about individuality; eliminativism about objects. Section \ref{sec:4} wraps it all up with metametaphysical considerations regarding the epistemology of metaphysics of science.\end{abstract}


\bodymatter

\section{Introduction}\label{sec:1}
The underdetermination of theory by data is a familiar topic in the general philosophy of science. It arises from the fact that there can be, in principle, more than one scientific theory that explains the same phenomena/data [\refcite{vanfraassen1980}]. At first, this kind of underdetermination was thought to be hypothetical at best [\refcite{stanford2021}], but nowadays it is well-established that quantum mechanics exemplifies it well [\refcite{callender2020}], viz. with the solutions to the measurement problem [\refcite{maudlin1995}, \refcite{arroyo-olegario2021}] e.g. Bohmian mechanics, Everettian quantum mechanics, and collapse-based quantum theories [\refcite{durrlazarovici2020}]. Metaphysical underdetermination is a further problem. It appears when a scientific theory is compatible with more than one metaphysical profile, and it is widely known that quantum mechanics {\it also} exemplifies it well with the discussion concerning the fact that quantum objects can be metaphysically understood both as individuals {\it and} as non-individuals [\refcite{french-krause2006}, \refcite{french1998}].

The first kind of underdetermination clearly supports an anti-realist argument against scientific realism. After all, if scientific realism is roughly defined as the stance according to which scientific theories are (in some sense)\footnote{~It is up for grabs whether `truth' should be understood in a correspondentist sense [\refcite{ladyman-ross2007}] or in a partial/quasi-truth sense [\refcite{dacosta-french2003}]. We shall not dwell on such matters here, however.} {\it true} descriptions of the world. Thus, in case of two (or more) competing theories to account for the same phenomena ---often positing different entities in its ontological catalog and different dynamics with different state spaces and axioms [\refcite{dacosta-bueno2011}, \refcite{ruetsche2018}, \refcite{arroyo-olegario2021}]--- it is not clear how to choose which one is true in a sense that is {\it of interest} for scientific realists. That is, the first kind of underdetermination prevents us from saying how the world looks like from the point of view of the ultimate furniture of the world.

When it comes to the second kind of underdetermination, however, it is not obviously true that anti-realism may reap any benefit specifically from it. In this chapter, we hope to clarify the lessons that the philosophy of science may learn from such kind of underdetermination. Section \ref{sec:2} presents the traditional debate over quantum individuality and the birth of the term ``metaphysical underdetermination''. Section \ref{sec:3} discusses the reactions to metaphysical underdetermination: one that employs extra-empirical virtues to favor the non-individuality metaphysical profile [\refcite{debarros-holik-krause2017}, \refcite{krause-arenhart2020}]; one that justifies the use of individuality [\refcite{muller2011}]; finally, the one that considers metaphysical underdetermination to be an argument {\it for} scientific realism [\refcite{french2014}, \refcite{french2020motivational}], as long as the realist content concerns structures, not objects [\refcite{ladyman1998}, \refcite{french-ladyman2003}, \refcite{french-ladyman2011}]. Finally, section \ref{sec:4} wraps it all up with epistemic considerations, viz. that once physics does not decide between metaphysical profiles, are we epistemically justified in adopting a philosophical attitude of (in)tolerance towards them [\refcite{carnap1950}]?

\section{Whence metaphysical underdetermination?}\label{sec:2}

Both classical and quantum particles are indistinguishable regarding state-independent properties (e.g., rest mass, electric charge). There is, however, a difference in how classical and quantum particles behave collectively. Such difference, captured by different statistical descriptions, is the source of the debate on quantum metaphysics that we address in this section. Since its earliest formulations, quantum-mechanical systems are known to obey the ``indistinguishability postulate (IP)'' defined as follows.

\begin{quote}
    Particles are indistinguishable if they satisfy the indistinguishability postulate (IP). This postulate states that all observables $O$ must commute with all particle permutations $P:[O,P]=0$.\footnote{~N.B.: here there is a typical abuse of notation. Officially, $[O,P]$ is an operator, not a scalar.} Put informally, the IP is the requirement that no expectation value of any observable is affected by particle permutations. \cite[p.~312]{huggett-imbo2009}.
\end{quote}

Let us consider the case of a system composed of $n$ indistinguishable particles, each particle $i \leq n$ having $\mathcal{H}_i$ as its state space. The state space of the whole system, described by $\mathcal{H}_n$, is written as: 
\begin{equation}
    \mathcal{H}_n \equiv \bigotimes^{n}_{i=1}\mathcal{H}_i
\end{equation}

Assume that $\mathcal{H}_n$ is closed under the permutation operator $P_{ij}$ which exchanges the $n$ factors of $\mathcal{H}_n$, i.e., $i$th and $j$th copies in $\mathcal{H}_n$. For $n=2$:
\begin{equation}
    P_{12}\big(|\varphi\rangle\otimes|\psi\rangle\big)=|\psi\rangle\otimes|\varphi\rangle
\end{equation}

Moreover, the IP says that the tensor product is not commutative so that $\langle \psi | \hat{A} | \psi \rangle = \langle P\psi | \hat{A} | P\psi \rangle$. The theoretical consequences of the IP are usually presented through the counting process for particles by comparing the Maxwell--Boltzmann (classical) statistics with the Bose--Einstein and the Fermi--Dirac (quantum) statistics, which enables one to visualize what is at stake. Suppose one wishes to arrange two particles, $A$ and $B$ in two boxes, $1$ and $2$, as depicted in table \ref{tab:statistics}.

\begin{table}[ht]
\centering
\begin{tabular}{lcc}
\hline\noalign{\smallskip}
& \textbf{Box $1$} & \textbf{Box $2$}\\
\noalign{\smallskip}\hline\noalign{\smallskip}
1.&$AB$ &     \\
2.& & $AB$    \\
3.&$A$ & $B$  \\
4.&$B$ & $A$  \\
\noalign{\smallskip}\hline
\end{tabular}
\caption{Statistics for particles in boxes}
\label{tab:statistics}
\end{table}

Within the Maxwell--Boltzmann statistics, there are four options to arrange the particles, to each of which we assign the equal probability weight of $\tfrac{1}{4}$. Regarding intuitiveness, so far so good. In the quantum case, due to IP, things are slightly different. Since quantum particles of the same kind are {\it indistinguishable}, one cannot assign any discerning role to the labels attributed to them ---such as ``$A$'' and ``$B$''--- in any physically meaningful way. While the first and second options resemble the Maxwell--Boltzmann probability weight of $\tfrac{1}{4}$ for each, the third and fourth options do not. Due to IP, one can permute particles $A$ and $B$ without changing the system's state; in the case of fermions, the state changes signal, but its square, which stands for the probability, is the same. So, instead of attributing a probability weight of $\tfrac{1}{4}$, we write:
\begin{equation}
    \frac{1}{\sqrt{2}}\Big(|A_1\rangle\otimes|B_2\rangle\pm|A_2\rangle\otimes|B_1\rangle\Big)
\end{equation}
for cases 3. and 4. That's the theoretical difference between classical and quantum statistics in a nutshell ---the addition being for bosons and the subtraction for fermions. To put it bluntly, what matters for our purposes is to state the fact that, from the experimental point of view, in the quantum case the permutation of particles does not give rise to a different physical situation. But in the classical case, it does: exchanging both particles $A$ and $B$ generates a different physical situation.

There are two standard ways of accommodating these matters in metaphysical terms regarding individuality. The first is to point out that in the classical realm, the difference in weighting probability in cases 3. and 4. implies that classical particles are individual objects in a metaphysical sense. Moreover, such individuality must be grounded in {\it something way beyond} the particles' state-independent properties, since they are indistinguishable with regard to those too. Even if they are absolutely indistinguishable, classical objects (e.g., $A$ and $B$) are typically regarded as individuals; that is, there is some metaphysical feature that accounts for the fact that $A$ is $A$, not $B$ and vice versa. This would explain the statistical difference between 3. and 4. Due to Post [\refcite{post1963}], this `something else' grounding individuality is known as ``transcendental individuality'' [\refcite{french1989}, \refcite{french-krause2006}], and it is typically framed in terms of an additional non-qualitative ingredient, like a substratum or an individual essence. Other options consist in attributing a unique space-time position to each particle and backing it with the Impenetrability Assumption, granting thus what is called `space-time individuality' \cite[chap.~2]{french-krause2006}. Quantum particles, by contrast, are objects that by virtue of assigning the same probability weight to cases 3. and 4. are said to have lost their individuality. To some founding fathers of quantum mechanics, such as Schr\"odinger \cite[p.~197]{schrodinger1998}, this was enough to state that quantum particles are non-individual objects [\refcite{weyl1931}]. That is because quantum objects obey the IP; the metaphysical lesson to be learned is that the metaphysics of individuality should be revised (or abandoned) thoroughly. This is the standpoint of the ``Received View'' on quantum individuality [\refcite{arenhart2017synt}], nowadays also called the ``orthodoxy'' [\refcite{bigaj2022}]: quantum mechanics forces us to abandon the universal application of our notion of metaphysical individuality.

The second way to accommodate the odd statistics of quantum-mechanical systems is to resist the revisionary metaphysical maneuver. As argued by French \cite[pp.~442--444]{french1989} (see also [\refcite{french-krause2006}]), quantum particles can {\it also} be seen as individuals, just like classical particles. In such a scenario, asymmetric states formed by permutations of particles' labels would not be counted not because they do not exist, but because they would not be available for the particles; such states are mathematically available, but not physically available, given that they violate the permutation symmetry required from quantum particles \cite[p.~148]{french-krause2006}. Quantum statistics, in this view, would differ from the classical case, not because of objects lacking individuality principles, but in terms of accessibility of states, which is restricted to certain subspaces $\mathcal{H}$ ---so the labels {\it can} reflect the individuality of quantum systems. In metaphysical terms, this individuality may be cashed out in terms of transcendental individuality, such as haecceity, primitive individuality, or by the substratum theory. As remarked by Caulton \cite[pp.~581--582]{caulton2022}, there is a strict link {\it between haecceitism and anti-haecceitism} and {\it transcendental individuality (TI) and qualitative individuality (QI)}, both defined as follows in permuted states $\mathfrak{W}$ and the states' common domain $D$:

\begin{quote}
({\it Haecceitism}) Any two distinct states permutable in $\mathfrak{W}$ represent distinct possibilities.\\({\it Anti-haecceitism}) Any two distinct states permutable in $\mathfrak{W}$ represent the same possibility. \newline[\dots]\newline (TI) Each label in $D$ denotes some object of the target system, and that label denotes the same object in all states.\\(QI) The labels in $D$ denote no object in particular of the target system. \cite[pp.~581--582, original emphasis]{caulton2022}
\end{quote}

What is more relevant for us now is the fact that quantum theory, by itself, does not provide any kind of support for any of the options. Such a situation concerning the metaphysics of quantum objects enables French and Krause \cite[p.~xiii, see also chap.~4]{french-krause2006} to state that ``[\dots]  there exists a form of `metaphysical underdetermination' between two conceptual packages which are both supported by the physics: particles-as-individuals and particles-as-non-individuals''. What is important to stress, again, is that quantum mechanics is in no position to decide between them, i.e. the metaphysics of individuality is underdetermined by quantum mechanics. Furthermore, from now on we will use the taxonomy offered by Krause and Arenhart [\refcite{krause-arenhart2018}], according to which there are three key concepts in this debate, each related to a distinct philosophical dimension: identity (a logical relation), individuation (a matter of epistemology), individuality (a non-relation property, a concern for metaphysics). So one could say, for the sake of an example, that quantum objects can be dealt with by a logical system with {\it identity}, even though they cannot be {\it discerned} or {\it individuated} in certain physical circumstances (e.g. when in Bose--Einstein condensates), and still maintain on the top of that a metaphysical profile of individuality or non-individuality. This is what concerns us here. As the taxonomy employed makes it possible to discuss each of these components (which are often confused or, at least, used interchangeably in the literature), the subject of this chapter is exclusively metaphysical, viz. on the question of individuality. As discussed in Krause and Arenhart [\refcite{krause-arenhart2018}], individuality, the metaphysical dimension, may be related to identity and discernibility, so that we have the qualitative versions of individuality, or else it may not be so related, in case we have versions of transcendental individuality.

This characteristic of the metaphysics of quantum mechanics generated at least three types of philosophical reactions, which range from considering it to be a feature or a bug. Some anti-realists consider metaphysical underdetermination to be a bug, so we should say ``good-bye to metaphysics'' \cite[p.~480]{vanfraassen1991}. Surely this is a stance one can voluntarily adopt [\refcite{chakravartty2017}]; one can even decide to be a realist about e.g. electrons but refuse to ask metaphysical questions about individuality [\refcite{psillos1999}]. This would put the scientific realist in the `shallow' spectrum of scientific realism [\refcite{magnus2012}, \refcite{french2018}, \refcite{arenhart-arroyo2021veri}]. The problem with this attitude is that without metaphysics, it is said, one loses sight of what one is being a realist about [\refcite{chakravartty2007}], thus decreasing our overall understanding of the theory [\refcite{french2014}]. One alternative to it is to point out that one can {\it also} voluntarily bite the bullet and keep asking metaphysical questions concerning the individuality profile of quantum objects ---a move which would put the scientific realist in the `deep' part of the spectrum [\refcite{magnus2012}, \refcite{french2018}, \refcite{arenhart-arroyo2021veri}]. The particular debate on deep realism for (non-)individuality in the context of quantum mechanics is grounded in the notion put forth by Brading and Skiles [\refcite{brading-skiles2012}] that objects must have their ``individuality profile'' spelled out, otherwise, we cannot make sense of objects whatsoever.

There are at least two realistic ways to proceed from there, which we'll cover in the next section.

\section{Three reactions}\label{sec:3}

The general message of metaphysical underdetermination for the philosophy of general science is this: as we are not able to empirically resolve the question of which metaphysical profile is the true one so that we can understand the nature of the entities with which scientific theories are committed, the question must appeal to non-empirical criteria.

Before doing that, however, one might be tempted to suggest leaving everything to the experimental part to decide. This is because the history of science has shown us that, through experiments, we can at least eliminate most of the possible metaphysics ---a great example would be the local metaphysics associated with standard quantum mechanics after Bell's theorems and Aspect's experiments [\refcite{aspect2002}]---, so it is not vindictive to say that we should completely rule out empirical criteria for metaphysical theory decisions. The difficulty of this suggestion is twofold: on the one hand, the experience is compatible with the metaphysical theories discussed here, viz. from particles-as-individuals and particles-as-non-individuals; on the other hand, if we adopt only what is authorized by experience, we end up with no metaphysics at all, with the constructive empiricists. And as we recall, the whole point was based on providing a realistic perspective on the metaphysics of quantum mechanics. Indeed, perhaps future experiments will falsify some of these metaphysical alternatives. But the point is that this is not the case now: the metaphysics of individuals, non-individuals, and structures are on an experimental equal footing. So we have two alternatives: 1) adopt ``quietism'' [\refcite{wolff2019}] towards metaphysics until we have a final and fundamental physical theory [\refcite{mckenzie2020}]; 2) opt for non-empirical criteria to decide on metaphysical theories. As a working hypothesis, we will assume the second.

Generally, such criteria are pragmatic, aesthetic, and/or metaphysical, such as simplicity (pragmatic), indispensability (metaphysics), beauty (aesthetic), and compatibility with other scientific theories (epistemic). In this section, we'll analyze the following realist reactions to metaphysical determination: to consider it to be a bug, so we have to break the underdetermination in order to be a realist about them (about the particles-as-non-individuals horn or the particles-as-individuals horn); to consider it to be a feature, so we have to embrace it to move to another form of realism.

\subsection{Non-individuality horn}\label{sec:3.1}

Let us recall the claims made by the Received View: quantum mechanics forces us to revise our metaphysical notions. Instead of building a notion of super-individuals, we are suggested to abandon it for good. One may take this idea literally, and benefiting from the fact that identity and individuality are generally brought together in these contexts, suggest a revision of the logic of identity [\refcite{arenhart2017synt}]. For instance, in classical logic, reflexivity of identity, viz. $x=x$ is trivially true, and if that is related to individuality, perhaps, failure of the latter requires failure of the former, no? But how can that be? As French and Krause stress: 

\begin{quote}
[\dots] the notion of non-individuality can be captured in the quantum context by formal systems in which self-identity is not always well-defined, so that the reflexive law of identity, namely, $\forall x(x = x)$, is not valid in general. \cite[p.~13--14]{french-krause2006}
\end{quote}

Logical systems in which the principle of reflexivity is relaxed are called ``non-reflexive logics'', which made, as French [\refcite{french2019sep}] put it, the notion of non-individuality ``philosophically respectable''; even more: as French \cite[p.~36]{french2014} emphasizes, ``[\dots]  without them, this metaphysical position ---of quantum particles as non-individuals--- might not be treated as a viable `horn' of an underdetermination argument at all''.

Briefly, some self-avowed theoretical (albeit non-empirical) virtues of this formulation of the Received View (viz. particles-as-non-individuals) are:

\begin{description}
\item[Simplicity] As it is well-known, the development of quasi-set theory [\refcite{krause1991}, \refcite{krause1992}, \refcite{french-krause2006}] enabled one to treat indistinguishability and non-individuality {\it right from the start} in quantum mechanics. As an alternative to standard non-relativistic quantum mechanics, non-reflexive quantum mechanics [\refcite{krause-arenhart2016}] employs quasi-set theory as the theory's logical foundation. Regarding mathematical simplicity, we should favor the Received View.

\item[Unity of science.] Ionization is a chemical process by which an atom acquires a negative charge by gaining electrons and acquires a positive charge by losing electrons. As Krause [\refcite{krause2019}, \refcite{krause2005}] emphasizes, chemistry wouldn't work properly with the assumption that the electron the atom loses is different from the electron the atom gains in the ionization process, viz. with the assumption that electrons are {\it individuals}. Regarding the unity of science, we should favor the Received View.

\item[Intuitiveness.] Due to the Kochen--Specker theorem, it is impossible to ascribe certain measurement/context-independent properties to quantum systems, which is a highly counter-intuitive result. As recently argued by de Barros, Holik, and Krause [\refcite{debarros-holik-krause2017}], one could attempt to argue that the Kochen--Specker theorem wouldn't hold if each measurement situation is considered to be indiscernible rather than identical, which is a feature that quasi-set theory and non-individuality can easily accommodate. So, regarding intuitiveness, we should favor the Received View.
\end{description}

\subsubsection{Logical underdetermination}\label{sec:3.1.2}

Regarding the abandonment of classical logic to cope with quantum non-individuality, Bigaj \cite[p.~42, fn.~24]{bigaj2022} states that ``[\dots]  quantum mechanics does not force us to adopt such a radical view''. To Bigaj,

\begin{quote}
    [\dots] quantum particles are definitely not individuals in the classical sense, [but] their non-individuality can hopefully be expressed without sacrificing classical logic that has served us well for two millennia. \cite[p.~245]{bigaj2022}
\end{quote}

    The upshot is that it would be nice if we stick with classical logics, so we wouldn't need to change logical systems {\it because of} the impositions of quantum mechanics. However, as shown by Arenhart \cite[p.~394]{arenhart2018}, {\it classical} systems with ``[\dots]  first- and second-order languages may be provided with unintended interpretations of identity which may do the job of failing identity and play the role of an indiscernibility relation''. So there's no {\it need} to change logics, strictly speaking, even if one defends the particles-as-non-individuals metaphysical package.

Bigaj's claim that quantum mechanics doesn't force us to adopt a non-classical logic can be considered on the following grounds: maybe he is right {\it if} the goal is just to get the probabilities of measurements, in a quite {\it instrumentalist} view in Bohr's [\refcite{bohr1963}] sense, that is, the use of classical logic and standard mathematics is compatible with the activities of the `practical' physicist; any book of QM you can find in your library is based on standard mathematics, hence in classical logic. But the problem can be viewed from another point of view, that of {\it foundations}, mainly {\it philosophical foundations}. Here the objectives are different, not that of the practitioner physicist, but of the philosopher interested in the meta-study of the theory itself, or better, of the cluster of theories we usually call ``quantum mechanics''. From this point of view, it is really fundamental to analyze the logical basis of the theory, and, as far as things go, apparently, the use of non-classical logics seems relevant and perhaps even necessary.

\subsection{Individuality Horn}\label{sec:3.2}

Quantum objects can also be cashed out as individuals, in metaphysical terms. As we already mentioned, one can metaphysically dress their individuality profile as individuals either as primitives [\refcite{morganti2015}] or in haecceistic terms: a quantum object has the quintessential/transcendental feature of being itself an individual. In metaphysical debates, this is known as the ``substratum theory'' of individuality, which states that individuality is a feature that obtains for objects over and above their properties. The substratum theory is an example of the Transcendental Individuality account. However, another account of individuality is available in textbook metaphysics: the ``bundle theory'' of individuality. In it, an individual is characterized {\it solely} by its properties. So, if two objects share all their properties, they have to be one and the same. This, of course, assumes the validity of Leibniz's Principle of the Identity of Indiscernibles (PII). However, as we mentioned above, quantum objects are indiscernible with regard to their state-independent properties. In a very rough reading, one could argue that quantum objects {\it refutes} the PII, so bundle theories are not options to metaphysically interpret quantum mechanics [\refcite{french-krause2006}]; alternatively, one could argue that, due to the PII, there is only one quantum object [\refcite{feynman1965}].

Somewhat recent development in the foundations of quantum mechanics showed, however, that the PII can be tailored for quantum-mechanical purposes. Quantum objects are said to be ``weakly discernible'' [\refcite{saunders2003}, \refcite{muller-saunders2008}, \refcite{muller2011}, \refcite{huggett-norton2013}]; that means that quantum entities may be discerned because entering into irreflexive and symmetric relations (and that is what `weakly discerning relations' mean). Given symmetry, a form of permutation is available: if $x$ is $R$ related to $y$, by symmetry, $y$ is $R$ related to $x$. Irreflexivity, on the other hand, accounts for the difference: if $R$ is a weakly discerning relation and $x$ is $R$ related to $y$, then, due to irreflexivity, it is not the case that $xRx$ and $yRy$, so that $x \neq y$. Such relations, then, are said to contribute to distinguishing two entities, saving a new version of the PII.

The traditional example concerns electrons' spin. Once electrons share the state-independent value of spin $\tfrac{1}{2}$, electrons cannot have an opposite spin value to themselves in a given axis. So, electrons $A$ and $B$ can be said to have opposite spin values from each other ---which is a way to discern them; electron $A$ has (say) opposite spin to electron $B$ in such a manner that $A$ and $B$ don't share the same properties anymore. So, as the argument goes, one can apply the PII to them. The conclusion is then that quantum mechanics is compatible with the PII, so bundle theory is indeed a metaphysical option to the individuality metaphysical profile of quantum objects.

On that note, Caulton [\refcite{caulton2013}] and Bigaj [\refcite{bigaj2021}] have recently argued that quantum particles are not indiscernible after all ---they are indeed absolutely discernible. One might argue ---in fact, it has been [\refcite{bigaj2022}]--- that this seems to be crucial for the metaphysical debate on quantum (non-)individuality, as it might be used to undermine the relevant form of underdetermination, viz. metaphysical underdetermination.

However, the {\it epistemic} notions of (in)discernibility do not need to be seen as determining the {\it metaphysical} issues concerning (non-)individuality. They can, but still, that connection itself is made in the discussion of the metaphysics; it does not come from quantum mechanics itself (see also [\refcite{krause-arenhart2020}]). As we apply the terminology, nothing ---except philosophical preferences--- determines the metaphysical profile of individuality or non-individuality.

The issue is similar to the remarks already advanced in the cases of weak discernibility for quantum particles. While one may attempt to connect this kind of discernibility with some approach to individuality directly related to discernibility, one is not so obliged by quantum theory; that is a decision of metaphysical nature. As Arenhart \cite[p.~110]{arenhart2017theo} puts it, ``[\dots]  there are many distinct and incompatible ways to put metaphysical flesh on the bare relational bones of weak discernibility''. That is, even if one considers that the obtaining of weakly discerning relations may be derived from the quantum mechanical formalism, the metaphysical lesson about individuality does not result obvious. Let us mention how that would work. One could hold, for instance, that due to what `individuality' actually means, weakly discerning relations are not enough to ground any kind of individuality worth its name. In a first approach, one could hold that individuality is conferred by some non-relational property. So, in the quantum case, although a weak version of PII could be seen as saved, individuality is still absent. For instance, due to the indistinguishability of quantum entities by their properties. In a second approach, one could still hold that individuality is granted by a transcendental principle of individuality, even though weak discernibility holds. In both cases, weakly discerning relations are not doing the expected metaphysical job. Also, in both cases, any metaphysical lesson drawn from weak discernibility must be added from outside of quantum mechanics. 

In this way, the particles-as-individuals horn has its own metaphysical underdetermination: are quantum objects individuals-as-substrata or individuals-as-bundles? Hence, {\it even if} one assumes that there are undisputed cases of empirical criteria favoring the metaphysical package of particles-as-individuals, there is still not a clear picture in sight for us to understand what is the metaphysical profile for quantum objects' individuality.

\subsection{Structuralist horn}\label{sec:3.3}

Recall that the `need' to ascribe a metaphysical profile concerning the (non-)individuality of quantum objects is to be {\it realist enough} about them. The claim, recall, is that without the metaphysical layer which floats free from what physics tells us about the nature of the entities with which it is ontologically committed to [\refcite{arroyoarenhart2022}], one cannot be considered to be realist enough about them. As French \cite[p.~50]{french2014} argues, without the metaphysical import of individuality or non-individuality for objects, one cannot understand what objects {\it are}, hence one cannot adopt a legitimate realist stance towards it. The metaphysical profile of individuality or non-individuality is, then, indispensable for object-oriented scientific theories [\refcite{brading-skiles2012}, \refcite{french2014}]. It was van Fraassen \cite[pp.~480--482]{vanfraassen1991} who first pointed out that the metaphysical underdetermination between individuality and non-individuality challenged the adoption of scientific realism; but as French \cite[p.~37]{french2014} argues that a particular form of scientific realism is challenged by metaphysical underdetermination: ``object-oriented realism''. What if one changes the ontological basis of scientific theories? Here's Ladyman [\refcite{ladyman1998}]:

\begin{quote}
It is an ersatz form of realism that recommends belief in the existence of entities that have such ambiguous metaphysical status. What is required is a shift to a different ontological basis altogether, one for which questions of individuality simply do not arise. \cite[pp.~419--420]{ladyman1998}
\end{quote}

In this passage, Ladyman [\refcite{ladyman1998}] uses the terms `ontology' and `metaphysics' interchangeably, which is not advisable insofar as it is a source of several problems in the metaphysics of science [\refcite{arenhart-arroyo2021manu}]. However, the point is that we should take metaphysical underdetermination as a ``motivational device'' to a dash of realism based on structures rather than objects, viz. structural realism [\refcite{french2011}, \refcite{french2014}, \refcite{french2020motivational}]. There are several forms of structural realism [\refcite{ladyman-2020}], our focus here being the ontological form of structural realism. But it wouldn't suffice to say that one believes in the structural components that remain through scientific changes; one must account for the metaphysical imports of what structures are. Otherwise, structuralists would remain in the shallow part of scientific realism, which is not advisable by their own standards of a genuine form of scientific realism [\refcite{french2014}]. One way to put it is to consider structures to be {\it fundamental} entities [\refcite{french2022}], to the point that objects are eliminable from an ontological point of view [\refcite{french2019shps}]. 

This, however, seems to be little informative about the metaphysical nature of structures, as Arenhart and Bueno [\refcite{arenhart-bueno2015}] stressed. On this point, structuralists argue that the literature is not being fair to them by asking about the nature of structures. For instance, French writes that:

\begin{quote}
    [\dots]  there's a certain asymmetry in the debate whereby structural realists are (constantly) asked `what is structure?' but their non-structural friends and colleagues are almost never required to give an answer to the corresponding question `what is an object?'. And of course it is not as if the answer to the latter is utterly straightforward. \cite[pp.~4--5]{french2020whatis}.
\end{quote}

Let us recall, however, that the search for the individuality profile is the search for a metaphysical characterization of what objects are. Metaphysical underdetermination just happens to block such a way, but surely the individuality profile is not exhaustive concerning the metaphysical nature of objects. One can ask e.g. about {\it contrast} and/or {\it extension} [\refcite{rettler-bailey2017}]. When concerning metaphysics ---or, as we are calling it here, the ``metaphysical profile''---, Chakravartty \cite[p.~12]{chakravartty2019} already recalled that ``[i]t is {\it always} possible to ask finer-grained questions [\dots] ''.

As Bianchi and Gianotti [\refcite{bianchi-gianotti2021}] emphasize, however, structuralists usually take structures to be the kind of entities metaphysically characterized extensionally: extrinsic properties, relations, symmetry groups, etc. To put it in another way, there is still metaphysical underdetermination concerning how should we understand `structures' properly in metaphysical terms [\refcite{french2020whatis}]. How's that different from object-oriented realism? It is unclear why we should favor structure-oriented realism over object-oriented realism {\it because of} metaphysical underdetermination, as both scientific-realist stances fall prey to metaphysical underdetermination.

\section{Lessons (un)learned: the elimination of metaphysics}\label{sec:4}
Given those discussions, the point is: as there are no empirical factors that can decide between these metaphysical views, all we have are pragmatic/aesthetic criteria for adopting one metaphysical profile or another. However, none of these solutions is final, as pragmatic values are not conducive to truth. Thus, what remains is to adopt an ``irenic'' attitude, à la Carnap, about such metaphysical proposals. As the famous ``Principle of Tolerance'' recommends:

\begin{quote}
    Let us grant to those who work in any special field of investigation the freedom to use any form of expression which seems useful to them; the work in the field will sooner or later lead to the elimination of those forms which have no useful function. \cite[p.~40]{carnap1950}.
\end{quote}

Thus, it seems that the maximum epistemic justification that one can have in the face of such metaphysical profiles is the acceptance of their empirical adequacy, and not the leading to the truth that such metaphysical profiles could supposedly have. To adopt such a stance, however, is something that constructive empiricists have always recommended.

But is that enough, given the purposes we started with?? It seems that a stance accommodates the problem, but does not solve it. One way of perceiving the discomfort of such empiricist accommodation is through the notion of ``understanding'', as used by neo-Pyrrhonists such as Bueno:

\begin{quote}
    We obtain an understanding of the various possibilities that are available to make sense of the issues under consideration and the insights such possibilities offer even if neo-Pyrrhonists are unable in the end to decide which of them (if any) is ultimately correct. \cite[p.~13]{bueno2021}.
\end{quote}

It is difficult to see how the multiplicity of metaphysical options, however, gives us understanding rather than misunderstanding. After all, having several incompatible options for understanding what science tells us about the world leaves us in a situation of \textit{confusion}, not enlightenment. In this way, perhaps, the biggest metametaphysical lesson that metaphysical underdetermination can bring us is the following: metaphysics, understood as an extra explanatory layer in relation to the scientific layer, must be avoided under the penalty of bringing more harm than gains.

Maybe van Fraassen [\refcite{vanfraassen1991}] was right after all: metaphysical underdetermination means good-bye to metaphysics ---or, at least, a farewell to its epistemic dignity for the time being. If metaphysics is too permissive from a methodological point of view, then it is easy to indicate the metaphysical profiles arbitrarily and to proliferate metaphysical profiles/options at any convenience. So we believe that ``good-bye'' is a farewell to the epistemic credentials of metaphysics ---something that scientific metaphysicians/naturalists were keen to provide [\refcite{ladyman-ross2007}, \refcite{bryant2020}, \refcite{hofweber2021}, \refcite{french2018jgps}]. That is, what is the epistemic value of a discipline in which (almost) anything goes? It seems to be very cheap. In any case, if the initial objective of justifying the introduction of a metaphysical layer of explanation was to increase our understanding of a certain domain of knowledge, well, it seems, the multiplicity of options ends up completely thwarting this ambition.

{\it This}, we think, is {\it the} most important metaphilosophical lesson unlearned from metaphysical underdetermination.

\section*{Acknowledgements}
Order of authorship is alphabetical and does not represent any kind of priority; authors have contributed equally to this chapter. A previous version of this work was presented at the Hi-Phi International Conference. Lisbon, 2022. We would like to thank Miguel Ohnesorge and Paulo Castro, whose constructive suggestions contributed to the improvement of this text. We would also like to thank Christian de Ronde for inviting us to participate in this volume, and for the always thought-provoking conversations about the philosophy of physics and scientific realism. Raoni Wohnrath Arroyo acknowledges the support of grants \#2021/11381-1 and \#2022/15992-8, S\~ao Paulo Research Foundation (FAPESP); Jonas R. Becker Arenhart and Décio Krause acknowledge the partial support of the National Council for Scientific and Technological Development (CNPq).

\end{document}